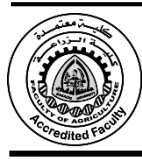
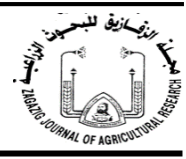

**Plant Production Science**

Available online at http://zjar.journals.ekb.eg
http:/www.journals.zu.edu.eg/journalDisplay.aspx?JournalId=1&queryType=Master

# EFFECT OF WATER SOAKING AND ROOTING SUBSTRATES ON THE ROOTING DEVELOPMENTOF SALAKHANI POMEGRANATE HARDWOOD CUTTINGS

**Iman J. B. Hawrami[1], M. M. A. Qadir[2] and R. R. Aziz[3]***

1. Nursing Dept., Darbanidikhan Techn. Inst. DTI, Sulaimani Polyt. Univ. SPU, Sulaimani, Kurdistan Region, Iraq

2. Garden Design Dept., Bakrajo Technical. Inst. BTI, Sulaimani Polyt. Univ. SPU, Sulaimani, Kurdistan Region, Iraq

3. Hort. Dept., Coll. Agric. Eng. Sci., Univ. Sulaimani, Kurdistan Region-Iraq

**ABSTRACT:** This study was carried out to evaluated the effects of soaking duration and rooting substrate on the root and shoot development of Salakhani pomegranate hardwood cuttings. The experiment was conducted at Darbandikhan Technical Institute, Sulaimani Polytechnic University, Iraq, from February to June 2025. The results showed that soaking duration had significant effect on rooting percentage, while the highest rooting percentage (60%) was obtained from the non-soaked cuttings, but prolonged soaking (48 h) resulted in the lowest rooting percentage (33.33%). Rooting substrate had a significant impact, with sand showing superior performance (75.56%) compared to the mixture (46.67%) and peat moss alone (20%). Moreover, the interactions between soaking duration and substrate revealed that soaking for 24 hours combined with sand yielded the highest rooting percentage (93.33%), while 48-hour soaking in peat moss led to complete rooting failure (0%). Root length and number were maximized in the 48 h + sand and 48 h + sand/peat moss treatments, while root biomass was highest in the 24 h + sand combination. Shoot characteristics were generally best in the control (0 h soaking), with the highest shoot length, diameter, weight, and chlorophyll content. However, the combination of 48 h soaking and sand/peat moss mixture resulted in the highest shoot length (35.44 cm) and leaf number (122), while 48 h + peat moss treatment suppressed all growth parameters. Consequently, the results suggest that immediate planting in sand or a sand/peat moss mix supports optimal rooting and shoot development.

## INTRODUCTION

The pomegranate (*Punica granatum* L.) belongs to the Punicaceae family and is the only species within the *Punica* genus cultivated for fruit production (**Salih *et al*., 2024**). Wild populations of pomegranate are extensively found across northern Iran, particularly along the Caspian Seashore and the Zagros Mountain plains, including areas such as Charmahal Bakhtiari, Fars, Kurdistan, Lorestan, Baluchistan, and the southern regions of the Alborz Mountain range. Due to this widespread presence, Persia is recognized as the center of origin for *Punica granatum* (**Zarie *et al*., 2021**). The, pomegranate is a deciduous shrub that generally grows to a height of 3 to 4 meters, although it can reach up to 9 meters under specific conditions (**Smith, 2014**). Also, this plant is well-known for being able to adapt and survive in dry, semi-dry, and salty environments. It is a good crop for places where other fruit trees have trouble with the weather (**Singh *et al*., 2011**). The, pomegranate plant is highly regarded and appreciated not only

---

\* **Corresponding author:** Tel. :+ 9647729969774

E-mail address: rasul.aziz@univsul.edu.iq

for its beautiful seeds but also for its health benefits. It is rich in antioxidants, especially polyphenolic compounds such as ellagic acid and punicalagin, as well as important vitamins, minerals, and tannins. The scientific study shows that these bioactive compounds may improve the immune system and overall health, which could lower the risk of heart disease and several cancers (**Sarrou *et al.*, 2014**). Pomegranate, peels and seeds are good for more than just their health benefits; they are also good for getting oils, pectin, and phenolic compounds. These byproducts also have notable industrial uses, acting as raw materials for the creation of biochar, biogas, and bio-oil (**El-Shamy and Farag, 2021**). Although, pomegranates can be propagated from seeds however this method is mostly used in breeding programs to create new cultivars because it changes the genetics of the plants and doesn't keep certain desirable qualities (**Prabhuling and Huchesh, 2018**). Vegetative propagation methods are better for making plants of the same quality every duration. Techniques like air layering, grafting, cuttings, and micropropagation aid in maintaining genetic homogeneity. Since hardwood cuttings can produce plants that are stronger and more vigorous than softwood and semi-hardwood cuttings, they are the most widely used of these. (**Saroj *et al.*, 2008**). However, successful plant propagation requires the selection of a suitable rooting medium. According to studies similarto those shown by **Giacobbo*et al.* (2007)**, quince cutting researchers have shown that the type of rooting medium used can have a substantial effect on the rooting success percentage. According to **Kumar *et al.* (2019)**, the optimal rooting environment should include a well-balanced mixture of water retention, aeration, nutrient availability, and structural stability in order to encourage vigorous root growth. Furthermore, **Aziz *et al.* (2020).** They found that soaking stem cuttings in water before planting significantly increases rooting success. This method allows the cuttings to go through advantageous metabolic changes by eliminating natural rooting inhibitors. Enhancing their capacity for rooting. This study aims to evaluate the effects of water soaking duration and rooting substrate on the success of Salakhani pomegranate (*Punica granatum*) hardwood cutting propagation.

## MATERIALS AND METHODS

This study was conducted to assess the effect of different rooting substrates and water soaking durations on the rooting performance of hardwood cuttings of Salakhani pomegranates (*Punica granatum* L.). The experiment was carried out from February 20 to June 20, 2025, at the Darbandikhan Technical Institute, Sulaimani Polytechnic University, Iraqs. Hardwood cuttings were collected on February 20, 2025, from three year-old Salakhani pomegranate trees grown in Halabja Governorate, Kurdistan Region. However, the cuttings were selected from the basal portion of one-year-old shoots and trimmed to a uniform length of 25 ± 1 cm and a diameter of 10 ± 2 mm to ensure consistency. The experiment was designed by using two factors: the first was water soaking duration at three levels (no soaking as control, soaking for 24 hours, and soaking for 48 hours), with the basal 5 cm of each cutting immersed in tap water for the designated duration; the second factor was the rooting substrate, which included (sand, peat moss, and a 1:1 volume-to-volume mixture of sand and peat moss). After soaking, the cuttings were planted in black plastic pots (28 × 26 cm) filled with the respective rooting media. The experimental layout followed a Randomized Complete Block Design (RCBD) under uncontrolled conditions (**Al-Rawi and Khalafullah, 2000**). A total of 135 cuttings were used, arranged across nine treatment combinations (three soaking durations × three substrates), with each treatment replicated three durations and five cuttings per replicate. The experiment continued until June 20, 2025, when data were collected on various rooting and shoot development traits. The recorded parameters included rooting percentage, number and length of main roots, root fresh and dry weights, number of shoots and leaves, shoot length and diameter, shoot fresh and dry weights, and total chlorophyll content. Data were statistically analyzed using XLSTAT software, and mean comparisons were conducted using Duncan's Multiple Range Test (DMRT) at a 5% significance level.

## RESULTS AND DISCUSSION

Fig. 1 illustrates that the effect of soaking duration on the rooting percentage of Salakhani pomegranate hardwood cuttings. Although the differences among soaking durations statistically significant, the highest rooting percentage (60%) was observed in the control treatment (0 hours), while the lowest (33.33%) occurred in cuttings soaked for 48 hours. Cuttings soaked for 24 hours exhibited an intermediate rooting percentage of 48.89%. These results suggest that extended soaking may adversely affect rooting success, with unsoaked cuttings producing the most favorable outcomes. Comparable findings were reported by **Salih et al. (2024)**, They demonstrated that excessive irrigation—functionally similar to continued soaking—led to reduced rooting percentages in certain pomegranate genotypes. In particular, applying irrigation daily (at 1-day intervals) significantly reduced rooting success, likely due to limited oxygen availability and a higher incidence of cutting rot caused by overly saturated media. These results highlight the significance of properly controlling moisture levels during propagation, then root development can be adversely affected by both excessive watering and long soaking.

The effect of different rooting substrates on the rooting percentage of Salakhani pomegranate hardwood cuttings is shown in Fig. 2. The results show a significant degree of variance between the treatments, with sand exhibiting the maximum rooting percentage (75.56%), followed by a combination of sand and peat moss (46.67%), and peat moss alone (20%) showing the minimum. Sand may perform better than other soils because of its greater drainage and aeration, which prevent waterlogging and increase oxygen availability—two factors that are good for root growth. On the other hand, peat moss tends to hold excess water, potentially leading to poor air circulation, root decay, or fungal problems. Furthermore, the intermediate performance of the sand–peat moss mixture likely stems from its balanced moisture retention and drainage characteristics. These results agree with the findings of **Netam et al. (2020)**, They reported that better rooting in pomegranate cuttings when grown in well-drained media such as sand.

Fig. 3 showed that the rooting percentage of Salakhani pomegranate hardwood cuttings was significantly affected by the combination of soaking duration and rooting substrate. Planting cuttings in sand after soaking them for 24 hours produced the maximum rooting percentage (93.33%), whereas planting cuttings in peat moss after soaking them for 48 hours produced the minimum rooting percentage (0.00%). Notably, cuttings planted in a sand and peat moss mixture without any soaking also demonstrated high rooting success (86.67%). These findings emphasize the critical role of both soaking duration and substrate selection in optimizing rooting outcomes. Sand offered superior aeration and drainage, particularly when combined with moderate soaking, promoting favorable conditions for root initiation. In contrast, peat moss alone likely retained excessive moisture, limiting oxygen availability and impeding root formation. These observations are consistent with the results of **Manila et al. (2017)**, They stated that the maximum rooting in vermiculite under mist conditions, although their study did not assess soaking duration. In summary, moderate pre-soaking combined with well-aerated media such as sand appears to significantly improve the rooting efficiency of pomegranate cuttings.

The data in Table 1 show that soaking duration had a significant influence on the shoot characteristics of Salakhani pomegranate hardwood cuttings. The control treatment (0 hours soaking) produced the best results across most parameters, including shoot length (22.74 cm), shoot diameter (1.97 mm), shoot fresh weight (10.36 g), shoot dry weight (5.60 g), and chlorophyll content (13.52 SPAD). This suggests that immediate planting without soaking preserved the physiological integrity of the cuttings, supporting better shoot growth and overall vigor. In contrast, soaking for 24 and 48 hours resulted in reduced shoot quality. Shoot diameter decreased to (1.39 mm and 0.88 mm), respectively, possibly due to tissue softening or hypoxic stress. Fresh and dry shoot weights also declined, likely from leaching of essential compounds or metabolic disturbance. Interestingly, the number of shoots was highest (6.44) in the 48-hour treatment, indicating that soaking might stimulate bud break but at the cost of shoot quality. The lowest

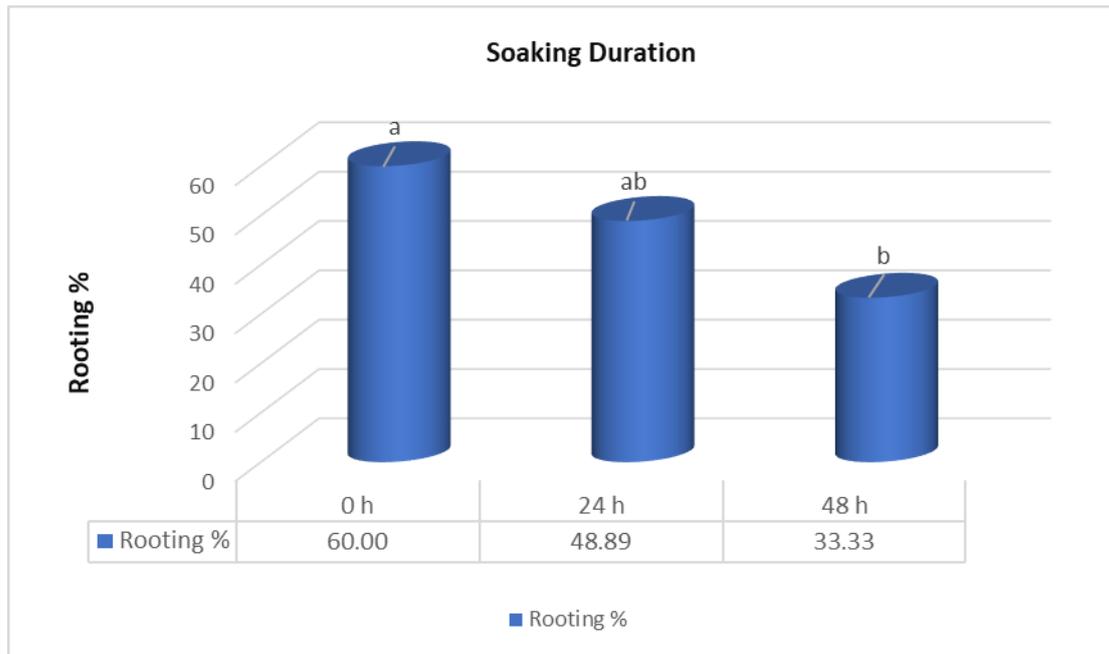

**Fig. 1.** Effect of soaking duration on the rooting percentage of Salakhani pomegranate hardwood cuttings

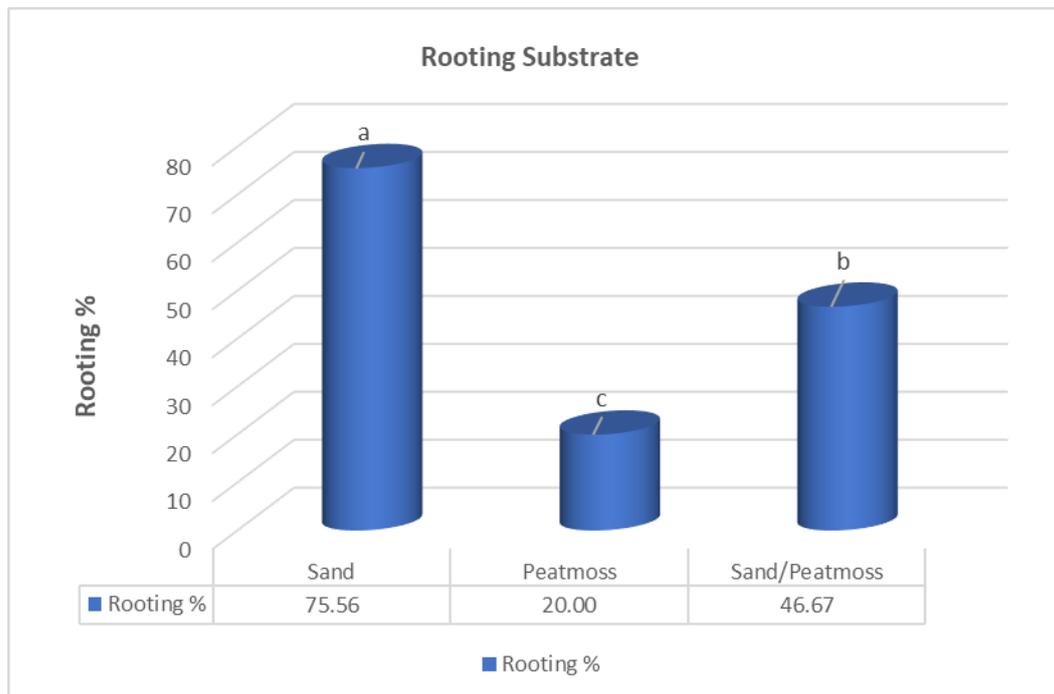

**Fig. 2.** Effect of rooting substrate on the rooting percentage of Salakhani pomegranate hardwood cuttings

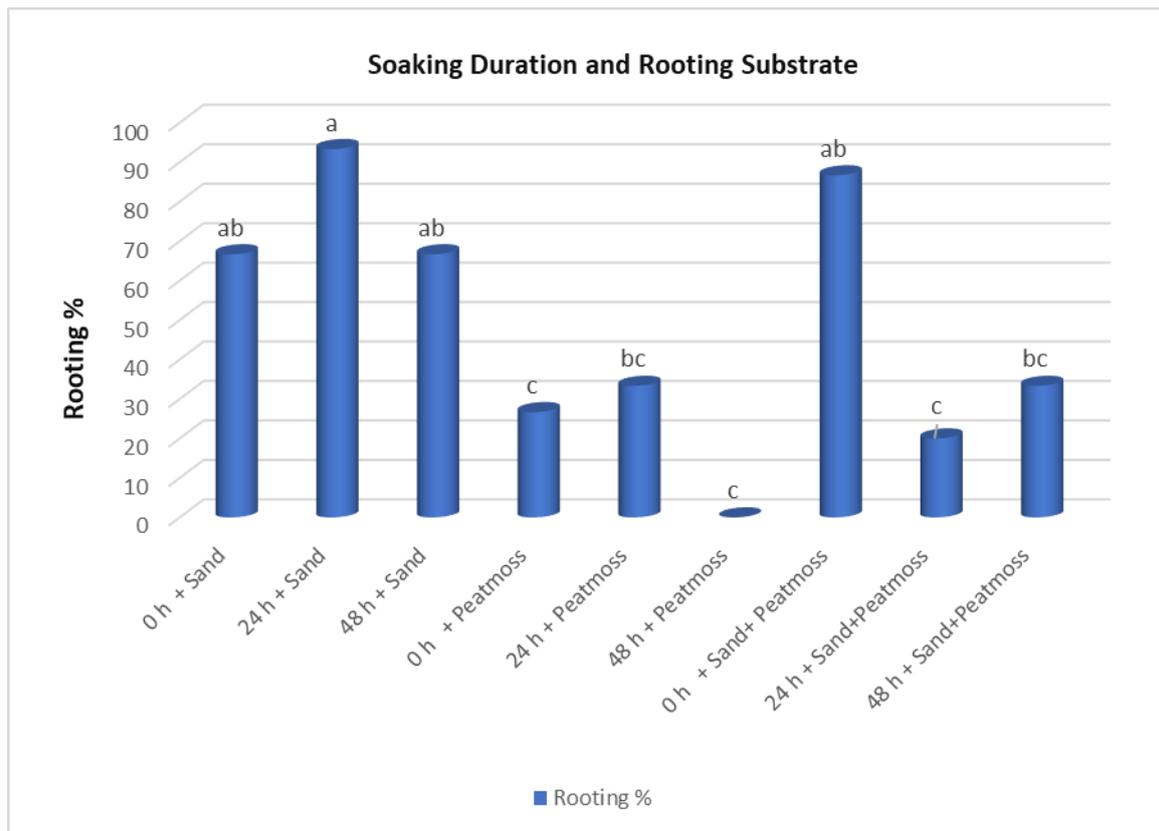

**Fig. 3. Interaction between soaking duration and rooting substrate on the rooting percentage of Salakhani pomegranate hardwood cuttings**

**Table 1. Effect of soaking duration on the shoot characteristics of Salakhani pomegranate hardwood cuttings**

| Soaking Duration (hours) | Number of shoots | Shoot length (cm) | Shoot diameter (mm) | Number of leaves | Shoot fresh weight (g) | Shoot dry weight (g) | Leaf chlorophyll content (SPAD) |
|---|---|---|---|---|---|---|---|
| **0 h** | 4.22 a | 22.74 a | 1.97 a | 81.00 a | 10.36 a | 5.60 a | 13.52 a |
| **24 h** | 3.52 a | 17.16 a | 1.39 b | 72.89 a | 8.67 ab | 4.62 a | 9.26 a |
| **48 h** | 6.44 a | 17.30 a | 0.88 b | 76.78 a | 6.34 b | 3.75 a | 12.40 a |

* Duncan's Multiple Range Test indicates that there is no significant difference (P≤0.05) between the values in any column that has the same letter.

chlorophyll content (9.26 SPAD) observed in the 24-hour soaking treatment may reflect stress-induced chlorophyll degradation. These observations are consistent with the findings of **Kurhe *et al*. (2022)**, they revealed that untreated or sub-optimally treated air layers showed reduced vegetative growth and survival, suggesting the importance of maintaining optimal pre-planting conditions to ensure shoot vigor.

Table 2 demonstrates that the effect of different rooting substrates (sand, peat moss, and a sand + peat moss mixture) on the shoot characteristics of Salakhani pomegranate hardwood cuttings. The data show that cuttings grown in sand and the sand + peat moss mixture generally outperformed those in peat moss alone. Although the number of shoots was highest in peat moss (5.48), this difference was not statistically significant compared to other treatments. However, shoot length, shoot diameter, number of leaves, fresh and dry shoot weight, and leaf chlorophyll content were significantly greater in cuttings grown in sand and the sand + peat moss mixture. For example, the longest shoots were observed in the sand + peat moss mixture (25.89 cm), followed by sand (21.00 cm), while peat moss produced considerably shorter shoots (10.31 cm). Similarly, cuttings cultivated in sand showed the highest shoot diameter, number of leaves, and chlorophyll content, indicating more vigorous shoot development. Because better drainage and aeration promote root activity and encourage shoot growth, sand and the sand–peat moss combo produced better outcomes. In contrast, peat moss may have limited root respiration due to its high water-holding capacity and restricted air exchange, which would impede shoot development. Enhancing shoot growth and physiological performance, the sand-peat moss combination seems to provide the best balance between air access and moisture retention among the tested media. These findings are consistent with previous studies by **Netam *et al*. (2020)**, they also reported that better rooting and shoot responses in sand-based or mixed media compared to peat moss alone. However, **Alikhani *et al*. (2011)** they showed peat-containing substrates some durations enhanced root formation, suggesting that the effect of rooting media may vary depending on species, cultivar, and environmental conditions.

Table 3 presents the interaction effects between soaking duration and rooting substrate on the shoot characteristics of Salakhani pomegranate hardwood cuttings. The results reveal that the combination of 48 hours of soaking and the peat moss + sand substrate produced the highest significant values for shoot length (35.44 cm), number of leaves (122.00), and leaf chlorophyll content (19.49 SPAD), demonstrating a strong positive influence on shoot development and physiological performance. Similarly, the highest shoot fresh weight (11.70 g) and shoot dry weight (6.77 g) were also recorded in this treatment. These results indicate that a well-balanced rooting substrate—combining the water retention of peat moss with the aeration properties of sand—supports optimal shoot growth under extended soaking conditions (**Bhagat and Dhaduk, 2009; Ghosh and Bera, 2005; Patel and Patel, 2001**). Interestingly, the highest number of shoots (10.00) was observed in the 48-hour soaking + peat moss treatment; however, all other associated shoot traits were zero. This suggests that these may have been undeveloped bud swellings or callus formations rather than true shoot development. In contrast, the same treatment (48-hour soaking + peat moss alone) recorded the lowest values for all shoot parameters, shoot length, diameter, leaf number, biomass, and chlorophyll content, with values of 0.00. This suggests that pure peat moss, due to its excessive moisture retention and poor aeration, led to severe suppression or failure of shoot development. Similar observations have been reported by **Islam *et al*. (2002)** found that the importance of proper aeration in rooting media to support shoot and root formation. These findings collectively suggest that extended soaking durations can be beneficial when combined with a well-drained substrate such as peat moss + sand, but may be detrimental when used with poorly aerated media like peat moss alone.

Table 4 shows the effect of soaking duration on the root characteristics of Salakhani pomegranate hardwood cuttings. Although the differences were not statistically significant (P ≤ 0.05), some trends were observed. The longest

**Table 2. Effect of rooting substrate on the shoot characteristics of Salakhani pomegranate hardwood cuttings**

| Rooting substrate | Number of shoots | Shoot length (cm) | Shoot diameter (mm) | Number of leaves | Shoot fresh weight (g) | Shoot dry weight (g) | Leaf chlorophyll content (SPAD) |
|---|---|---|---|---|---|---|---|
| Sand | 4.81 a | 21.00 a | 1.83 a | 103.11 a | 9.57 a | 5.41 a | 15.28 a |
| Peat moss | 5.48 a | 10.31 b | 0.92 b | 34.33 b | 5.71 b | 2.87 b | 6.60 b |
| Peat moss+ Sand | 3.89 a | 25.89 a | 1.49 a | 93.22 a | 10.09 a | 5.69 a | 13.31 a |

* Duncan's Multiple Range Test indicates that there is no significant difference (P≤0.05) between the values in any column that has the same letter.

**Table 3. Interaction between soaking duration and rooting substrate on the shoot characteristics of Salakhani pomegranate hardwood cuttings**

| Soaking Duration (hours) | Rooting substrate | Number of shoots | Shoot length (cm) | Shoot diameter (mm) | Number of leaves | Shoot fresh weight (g) | Shoot dry weight (g) | Leaf chlorophyll content (SPAD) |
|---|---|---|---|---|---|---|---|---|
| 0 h | Sand | 4.33 a | 19.000 bc | 2.24 a | 81.00 abc | 10.25 a | 5.41 a | 15.33 ab |
| 0 h | Peat moss | 3.99 a | 18.000 bc | 1.48 ab | 42.00 cd | 9.14 a | 4.62 a | 12.40 ab |
| 0 h | Sand +Peat moss | 4.33 a | 31.217 a | 2.20 a | 120.00 a | 11.70a | 6.77 a | 12.84 ab |
| 24 h | Sand | 4.55 a | 27.550 ab | 1.95 ab | 120.00 a | 10.47 a | 5.92 a | 12.79 ab |
| 24 h | Peat moss | 2.44 a | 12.943 c | 1.29 ab | 61.00 bc | 8.00a | 3.98 a | 7.40 b |
| 24 h | Sand +Peat moss | 3.55 a | 10.997 c | 0.93bc | 37.67 cd | 7.55 a | 3.96 a | 7.60 b |
| 48 h | Sand | 5.553 a | 16.443 bc | 1.31 ab | 108.33 ab | 7.98 a | 4.90 a | 17.71 a |
| 48 h | Peat moss | 10.00 a | 0.000 d | 0.00 c | 0.00 d | 0.00 b | 0.00 b | 0.00 c |
| 48 h | Sand +Peat moss | 3.77 a | 35.443 a | 1.34 ab | 122.00 a | 11.02 a | 6.35 a | 19.49 a |

* Duncan's Multiple Range Test indicates that there is no significant difference (P≤0.05) between the values in any column that has the same letter.

**Table 4. Effect of soaking duration on the root characteristics of Salakhani pomegranate hardwood cuttings**

| Soaking Duration (hours) | Root length (cm) | Number of roots. | Root fresh weight (g) | Root dry weight (g) |
|---|---|---|---|---|
| 0 h | 21.44 a | 35.22 a | 10.04 a | 4.84 a |
| 24 h | 24.11 a | 40.11 a | 6.62 b | 3.84 a |
| 48 h | 25.22 a | 38.56 a | 6.34 b | 3.48 a |

* Duncan's Multiple Range Test indicates that there is no significant difference (P≤0.05) between the values in any column that has the same letter.

average root length (25.22 cm) and highest number of roots (40.11) were recorded in the 48-hour and 24-hour soaking treatments, suggesting that soaking may enhance root initiation and elongation (**Nawaz and Khan, 2010; Ahmed and Hassan, 2015**). However, the control treatment (0 hours soaking) produced the greatest root fresh weight (10.04 g) and dry weight (4.84g), While, soaking improved root length and number to some extent, longer soaking durations resulted in lower root fresh and dry weights, possibly due to early water absorption reducing the stimulatory effect of the rooting substrate (**Nawaz and Khan, 2010**). Conversely, some studies have reported that longer soaking durations can increase root biomass along with root length and number (**Singh and Singh, 2012; Khan and Rashid, 2013; Patel and Patel, 2010**), which contrasts with the present findings. These differing results suggest that the effect of soaking duration on root development may vary depending on species, cutting type, and environmental conditions.

Table 5 presents the effect of different rooting substrates on the root characteristics of Salakhani pomegranate hardwood cuttings. The peat moss + sand mixture produced the longest roots (29.89 cm), closely followed by sand alone (27.22 cm), with no significant difference between these two substrates. Both sand and the sand + peat moss mixture also yielded significantly higher root numbers (50.11 and 44.67, respectively) compared to peat moss alone (19.11). The highest root fresh weight (11.24 g) and dry weight (6.34 g) were observed in the sand substrate, reflecting its superior aeration and drainage that favor vigorous root biomass development. Conversely, peat moss alone resulted in significantly lower values for all root parameters, likely due to excessive moisture retention and poor aeration, reducing oxygen availability and root function. These results agree with findings by **Patel and Patel (2001)** and **Nawaz and Khan (2010)**, who reported enhanced rooting in sandy or mixed substrates due to better aeration. Conversely, **Islam *et al*. (2002)** highlighted that peat moss alone can limit root growth because of reduced oxygen diffusion. Thus, sandy substrates, especially when combined with peat moss, provide an optimal balance for root growth in pomegranate cuttings.

Table 6 shows the interaction effects of soaking duration and rooting substrate on the root characteristics of Salakhani pomegranate hardwood cuttings. The best rooting performance was observed in the 48-hour soaking + sand treatment, which produced the highest values for root length (38.00 cm), number of roots (60.67), and root fresh (11.45 g) and dry weight (6.58 g), although these were not significantly different from other high-performing treatments. Similarly, the 48-hour soaking + peat moss + sand treatment demonstrated excellent rooting traits with a root length of 37.67 cm and 55 roots, highlighting the positive role of mixed substrates that balance moisture retention and aeration (**Patel and Patel, 2001; Nawaz and Khan, 2010**). In contrast, the 48-hour soaking + peat moss treatment resulted in complete failure, with zero values for all root parameters, likely caused by excessive moisture retention and oxygen deficiency inherent in pure peat moss (**Islam *et al*., 2002**). Notably, the highest root fresh weight (12.39 g) and dry weight (7.88 g) were recorded in the 24-hour soaking + sand treatment, emphasizing that moderate soaking combined with a well-drained substrate optimizes root biomass accumulation (**Singh and Singh, 2012**). Conversely, lower root weights were observed in the 24-hour soaking + peat moss and 24-hour soaking + peat moss + sand treatments, further indicating that peat moss alone or in excess moisture conditions impairs root development. Overall, these results suggest that soaking durations of 4 to 8 hours combined with sandy or mixed substrates promote the best rooting performance, while pure peat moss becomes detrimental when prolonged soaking is applied.

## Conclusion

This study concluded that both soaking duration and rooting substrate play a crucial role in the rooting and shoot development of Salakhani pomegranate (*Punica granatum* L.) hardwood cuttings. However, the best results were obtained with 24-hour soaking in sand, which provided the highest rooting percentage (93.33%). Sand proved to be the most suitable substrate due to its excellent drainage and aeration. In contrast, peat moss alone, especially under prolonged soaking (48 h), significantly reduced rooting and shoot growth. Overall, for successful propagation, cuttings should be planted in sand or a sand–peat moss mix, with soaking limited to 24 hours or avoided altogether.

Table 5. Effect of rooting substrate on the root characteristics of Salakhani pomegranate hardwood cuttings

| Rooting substrate | Root length (cm) | Number of roots. | Root fresh weight (g) | Root dry weight (g) |
|---|---|---|---|---|
| Sand | 27.22 a | 50.11 a | 11.24 a | 6.34 a |
| Peat moss | 13.67 b | 19.11 b | 5.15 b | 2.65 b |
| Peat moss+ Sand | 29.89 a | 44.67 a | 6.62 b | 3.17 b |

\* Duncan's Multiple Range Test indicates that there is no significant difference (P≤0.05) between the values in any column that has the same letter.

Table 6. Interaction between soaking duration and rooting substrate on the root characteristics of Salakhani pomegranate hardwood cuttings

| Soaking Duration (hours) | Rooting substrate | Root length (cm) | Number of roots | Root fresh weight (g) | Root dry weight (g) |
|---|---|---|---|---|---|
|      | Sand | 18.00 ab | 39.67 ab | 9.87 ab | 4.54 bc |
| 0 h  | Peat moss | 18.00 ab | 23.00 bc | 11.68 a | 5.95 abc |
|      | Peat moss+ Sand | 28.33 a | 43.00 ab | 8.58 ab | 4.02 cd |
|      | Sand | 25.67 a | 50.00 ab | 12.39 a | 7.88 a |
| 24 h | Peat moss | 23.00 a | 34.33 ab | 3.78 c | 2.00 de |
|      | Peat moss+ Sand | 23.67 a | 36.00 ab | 3.70 c | 1.64 de |
|      | Sand | 38.00 a | 60.67 a | 11.45 a | 6.58 ab |
| 48 h | Peat moss | 0.00 b | 0.00 c | 0.00 d | 0.00 e |
|      | Peat moss+ Sand | 37.67 a | 55.00 a | 7.56 b | 3.85 cd |

\* Duncan's Multiple Range Test indicates that there is no significant difference (P≤0.05) between the values in any column that has the same letter.

## REFERENCES

Ahmed, M., and Hassan, Z. (2015). Effects of soaking duration on rooting of pomegranate cuttings (*Punica granatum* L.). International Journal of Agriculture and Forestry, 5(2): 103-109.

Alikhani, L., Ansari, K., Jamnezhad, M., and Tabatabaie, Z. (2011). The effect of different mediums and cuttings on growth and rooting of pomegranate cuttings. *Iranian J. Plant Physiol.*, 1(3): 199-203.

Al-Rawi, K. M. andKhalafullah, A. (2000). Design and Analysis of Agricultural Experiments. Dar Al-Kutub for Printing and Publishing, University of Mosul. Ministry of Higher Education and Scientific Research, Iraq

Aziz, R.R., Mohammed, A.A., Ahmad, F.K., and Ali, A.J. (2020). Effect of IBA concentration and water soaking on rooting hardwood cuttings of black mulberry (*Morus nigra* L.). *Journal of Zankoy Sulaimani*. Part-A-Volume 22, Issue 1.

Bhagat, R. K., andDhaduk, B. K. (2009). Effect of different rooting media and concentrations of IBA on rooting of pomegranate (*Punica granatum* L.) cuttings. *The Asian Journal of Horticulture*, 4(1): 222–224.

El-Shamy, S., and Farag, M. A. (2021). Novel trends in extraction and optimization methods of bioactives recovery from pomegranate fruit biowastes: Valorization purposes for industrial applications. *Food chemistry*, 365, 130465. https://doi.org/10.1016/j.foodchem.2021.130465

Ghosh, S. N., and Bera, B. (2005). Effect of IBA, growing media and covering material on the rooting and sprouting of pomegranate cuttings (*Punica granatum* L.). *Environment and Ecology*, 23(3): 635–638.

Giacobbo, C.L., Fachinello, J. C. and Bianchi, V. J. (2007). Effect of substrate, indolebutyric acid and root grafting on the propagation of quince (*Cydonia oblonga* Mill.) cultivar EMC by cuttings. Ciência e Agrotecnologia, 31(1): 64-70.

Islam, M. N., Mollah, M. F. A., and Rahman, M. M. (2002). Rooting potential of cuttings of guava as influenced by rooting media and different concentrations of IBA. *Pakistan Journal of Biological Sciences*, 5(5): 562–564.

Khan, M. A., and Rashid, A. (2013). Influence of water soaking on root initiation and growth of hardwood cuttings of *Moringa oleifera*. *African Journal of Plant Science*, 7(5): 176-181.

Kumar, S., Malik, A., Yadav, R., and Yadav, G. (2019). Role of different rooting media and auxins for rooting in floricultural crops: A review. *International Journal of Chemical Studies*, 7(2): 1778-1783.

Kurhe, A. R., Hota, D., Verma, R., and Karna, A. K. (2022). Influence of different type of rooting media on rooting, growth and development of air layering in pomegranate. *International Journal of Horticulture and Food Science*. 4(2): 43-46

Manila, Tanuja, Rana DK, Naithani DC. (2017). Effect of different growing media on vegetative growth and rooting in Pomegranate (*Punica granatum* L.) Cv. "kandhari" hardwood stem cutting under mist. *Plant Archives*.17(1):391-394.

Nawaz, S., and Khan, M. A. (2010). Soaking treatments improve root initiation and length but do not always increase root biomass in cuttings of woody plants. *Pakistan Journal of Botany*, 42(3): 1939-1947.

Netam, SR., Sahu, GD., Markam, PS., and Minz, AP. (2020). Effect of different growing media on rooting and survival percentage of pomegranate (*Punica granatum* L.) cuttings cv. Super Bhagwa under Chhattisgarh plains condition. *International Journal of Chemical Studies*. 8(5):1517-1519.

Patel, N., and Patel, R. (2010). Impact of soaking duration on rooting of pomegranate cuttings under mist conditions. *Indian Journal of Horticulture*, 67(2), 245-248.

Patel, R. M., & Patel, N. M. (2001). Effect of different media and growth regulators on rooting of pomegranate (*Punica granatum* L.) hardwood cuttings. *South Indian Horticulture*, 49: 160–163.

Prabhuling, G., and Huchesh, H. (2018). In Vitro Regeneration in Pomegranate (*Punica granatum* L.) cv. Bhagwa using Double Nodal Segments. *Research Journal of Biotechnology*, 13(8): 1-10. https://ssrn.com/abstract=3080795

Salih, K. O., Mohammed, A. A., Faraj, J. M., Raouf, A. M., and Tahir, N. A. R. (2024). Rooting behavior of pomegranate (*Punica granatum* L.) hardwood cuttings in relation to genotype and irrigation frequency. *Journal of Agriculture and Environment for International Development*.118(1): 19 – 30. DOI: 10.36253/jaeid-13837

Saroj, P. L., Awasthi, O. P., Bhargava, R., and Singh, U. V. (2008). Standardization of pomegranate propagation by cutting under mist system in hot arid region. *Indian Journal of Horticulture*, 65(1), 25-30

Sarrou, E., Therios, I., and Dimassi-Theriou, K. (2014). Melatonin and other factors that promote rooting and sprouting of shoot cuttings in *Punica granatum* cv. Wonderful. *Turkish Journal*. https://doi.org/10.3906/bot-1302-55

Singh, B., Singh, S., and Singh, G. (2011). Influence of planting duration and IBA on rooting and growth of pomegranate (*Punica granatum* L.) 'Ganesh' cuttings. *Acta horticulture*, 890: 183-188. https://doi.org/10.17660/ActaHortic.2011.890.24

Singh, R., and Singh, M. (2012). Effect of soaking duration on rooting and growth of pomegranate cuttings. *Journal of Horticultural Science*, 7(1), 45-50.

Smith, R.E. (2014). Pomegranate Botany, Postharvest Treatment Biochemical Composition and Health Effects. Nova Science Publishers, USA

Zarie, A., Zamani, Z., and Sarkhosh, A. (2021). Biodiversity, germplasm resources and breeding methods. In A. Sarkhosh, A. M. Yavari, Z. Zamani (Ed.), The Pomegranate: Botany, Production and Uses (pp. 94-157). CABI, UK: Oxfordshire.

## تأثير نقع الماء ووسط التجذير في نمو الجذور وتطورها لأقلام الخشبية الصلبة لرمان صنف سالخاني

ايمان جزا باقى هورامى1، مشخل محمد امين قادر2، رسول رفيق عزيز3

1. قسم التمريض، المعهد التقنى دربنديخان، جامعة بوليتكنك السليمانية، السليمانية، إقليم كردستان، العراق.
2. قسم تصميم الحدائق، المعهد التقنى بكرجو، جامعة بوليتكنك السليمانية، السليمانية، إقليم كردستان، العراق.
3. قسم البستنة، كلية علوم الهندسة الزراعية، جامعة السليمانية، إقليم كردستان – العراق.

**الخلاصة**

اجريت هذه الدراسة لتقييم تأثير مدة النقع و نوع الوسط الزراعي على تجذير وتطور النمو الخضري لعقل الرمان الخشبي صنف "سالخاني". نفذت التجربة في معهد دربنديخان التقني، جامعة السليمانية التقنية، العراق، خلال الفترة من شباط الى حزيران 2025، باستخدام تصميم القطاعات العشوائية الكاملة بثلاث مكررات لكل معاملة. أظهرت النتائج ان مدة النقع لها تأثير معنوي في نسبة التجذير، حيث سجلت أعلى نسبة التجذير (60%) في العقل غير المنقوعة، بينما ادت فترة النقع الطويلة (48 ساعة) الى أدنى نسبة التجذير (33.33%). في المقابل، وأظهر نوع الوسط الزراعي تأثيراً معنوياً واضحاً، وسجلت اعلى نسبة التجذير (75.56%) اذا استخدام الوسط الرمل مقارنة بالخليط الرمل مع البيتموس (46.67%) والبيتموس منفرداً (20%). أما التداخل بين مدة النقع ونوع الوسط الزراعي، فقد أظهر أن النقع لمدة 24 ساعة مع الوسط الرملي سجلت اعلى نسبة التجذير (93.33%)، في حين أن النقع لمدة 48 ساعة داخل الوسط البيتموس أدى الى فشل كامل في التجذير (0%). كما سجل أطول طول للجذور وأكبر عدد منها في معاملتي 48 (ساعة + الرمل) و 48 (ساعة + خليط الرمل مع البيتموس)، بينما تحققت اعلى كتلة جذرية في معاملة (24 ساعة + الرمل). بالنسبة للصفات الخضرية، فقد سجل أفضل القيم في معاملة الشاهد لمدة (0 ساعة نقع)، حيث تحققت أعلى قيم لطول الساق، وقطره، ووزنه الطري، ومحتوى الكلوروفيل من الأوراق. ومع ذلك، فان معاملة النقع لمدة 48 ساعة مع خليط الرمل/البيتموس أعطت أطول ساق (35.44 سم) و أعلى عدد أوراق (122)، بينما أدى النقع لمدة 48 ساعة في البيتموس الى تثبيط جميع الصفات الخضرية. تشير النتائج الى أن الزراعة المباشرة في الرمل أو خليط الرمل/البيتموس توفر أفضل الظروف للتجذير و تطور النمو الخضري، في حين أن النقع لفترات طويلة، خاصة في الأوساط ضعيفة التهوية مثل البيتموس، يؤثر سلباً على أداء العقل.

**الكلمات الإسترشادية:** الإكثار بالعقل، النقع بالماء، وسط التجذير، تطوير البراعم.